\documentclass[aps,prl,twocolumn,floatfix]{revtex4}
\usepackage[dvips]{graphicx}

\begin{document}
\title{\boldmath The role of intra-atomic non-collinear
  magnetization density in weak ferromagnetism}

\author{Robert Laskowski}
\author{Gilles Santi}
\affiliation{Department of Physics and Astronomy, University of Aarhus,
 DK-8000 Aarhus C, Denmark}

 \date{\today}

 \begin{abstract}

 We investigate the mechanism behind the breakdown of the compensation of
 large magnetic moments leading to weak ferromagnetism. For this we use
 first-principles calculations within density functional theory and we
 focus on the weak ferromagnetic compound Mn$_3$Sn. Our new implementation
 allows for an exact treatment of the spin-density matrix and
 non-collinearity. In order to gain some insight, our results are compared
 to the ones obtained by using the atomic moment approximation (AMA) and
 its role is discussed. We find that the appearance of the weak
 ferromagnetic moment in this compound originates not so much as an effect
 of spin-orbit coupling as suggested previously from AMA calculations, as
 from the non-collinearity of the Sn atom magnetization density. This is
 confirmed by non-collinear calculations in which the SOC effects are
 neglected.

 \end{abstract}


 \maketitle

 Weak ferromagnetism is a phenomenom commonly characterized by a small net
 magnetic moment in a system of large localized moments nearly canceling
 each other. It has been known for more then 50 years and observed both in
 metallic and insulating materials (e.g. $\alpha$-Fe$_2$O$_3$, carbonates
 of Mn and Co, NiF$_2$, Mn$_3$Sn or Mn$_3$Ge). The earliest trial to devise
 a model Hamiltonian describing the mechanism behind it was put forward by
 Dzialoshinski \cite{Dzialoshinski1958} and was based on Landau's theory of
 second order phase transitions.  Moriya presented a more rigorous
 derivation for magnetic insulators \cite{Moriya1960}, extending Anderson's
 theory of superexchange to include spin-orbit coupling (SOC). Kataoka {\it
 et al.}  refined this theory further and applied it to metals
 \cite{Kataoka1984}. These models identify two types of interactions
 responsible for the appearance of weak ferromagnetism, i.e. the
 magneto-crystalline anisotropy and the antisymmetrical part of the
 anisotropic superexchange.

 This problem has also been approached from the point of view of
 first-principles calculations within density functional theory (DFT). An
 early local spin density approximation (LSDA) calculation for Mn$_3$Sn
 within the atomic moment approximation (AMA), which neglected SOC, was
 however not able to resolve the weak ferromagnetic structure
 \cite{Sticht1989}. Subsequently, including SOC effects into the DFT
 Hamiltonian, Sandratskii and K\"ubler concluded that this is essential to
 the correct description of weak ferromagnetism \cite{Sandratskii1996,
 Sandratskii1998}.

 In this Letter, we use our newly developed method \cite{Laskowski2004}
 that incorporates non-collinear effects into the all electron linearized
 augmented plane wave plus local orbital (L/APW+lo) WIEN2k code
 \cite{wien2k}.  Our implementation is based on a mixed spinor basis set
 approach \cite{yamagamincm,Kurz2004}. In the interstitial region the basis
 functions are pure spinors given in a global spin coordinate frame,
 whereas, inside the atomic spheres, they are non-pure spinors given in the
 (local) sphere's spin coordinate frame with the quantization axis pointing
 along the sphere's average magnetization. One advantage of this method is
 that it allows the use of spin-polarized radial basis functions. In
 addition, we have extended the implementation of Kurz {\it et al.}
 \cite{Kurz2004} described recently in order to treat the spins fully
 non-collinearly also inside the spheres. This means that the off-diagonal
 terms of the spin-potential matrix inside the atomic spheres are taken
 into account. The full spin-potential matrix is evaluated from the
 spin-density matrix using the methodology applied to interstitial charge
 \cite{Kurz2004}. The method uses the fact that the exchange potential is
 local and can be evaluated independently at each point of the unit cell.
 Because the standard LSDA formulation applies only to the collinear case
 (i.e. diagonal spin-density matrix) we rotate the spin-density matrix in
 spin-space to its diagonal form at each point in the unit cell
 independently. The diagonal potential matrix (at this point in space) is
 then evaluated in this ``local'' spin frame and rotated back into the
 original spin frame, leading to off-diagonal matrix elements.  As the
 treatment of non-collinearity require the ``full'' spin Hamiltonian, SOC
 \cite{soc} is naturally added to the scalar relativistic Hamiltonian in
 this first variational step.  To demonstrate the importance of such an
 exact treatment of the spin-density matrix, we focus, in this paper, on
 the case of the weak ferromagnet Mn$_3$Sn.

 \begin{figure}
      \includegraphics[width=\linewidth]{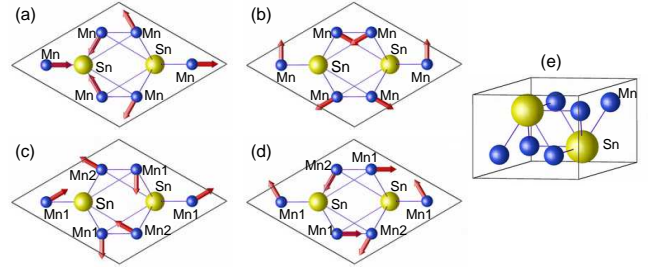}
 \caption{\label{structures} Candidate magnetic
    structures for Mn$_3$Sn (a)--(d) and unit cell (e). (The 4 Mn1 atoms and
    the 2 Mn2 ones are not equivalent in structures (c) and (d)).
 }
 \end{figure}

 The magnetic behavior of Mn$_3$Sn has been the subject of a number of
 experimental studies beginning from the early sixties. Weak ferromagnetism
 in this compound was first discovered by Yasukochi {\it el al.}
 \cite{Yasukochi1961} and Ohoyama \cite{Ohoyama1961}. According to several
 neutron diffraction measurements \cite[and references
 therein]{Tomiyoshi1982, Tomiyoshi1986, Tomiyoshi1987, Ohmori1987,
 Brown1990}, Mn$_3$Sn displays a quite interesting temperature dependence of
 the magnetization which depends however on the stoichiometry (the structure
 can be stabilized only in the presence of some excess of Mn) and the
 thermal treatment. It can be characterized by three temperatures: $T_N
 \approx 420$~K, at which weak ferromagnetism appears, $T_1 \approx$
 150--270~K, at which magnetization vanishes almost entirely (with the
 possible appearance of an incommensurate spin-spiral state), and $T_2
 \approx$ 50--100~K, at which the magnetization increases again to reach its
 maximum value for $T=0$. The chemical unit cell is of DO$_{19}$ type, with
 hexagonal space group $P6_{3}/mmc$, and contains two formula units. It is
 composed of two layers of Mn atoms arranged in triangles, with an Sn atom
 on top of each triangle (Fig. \ref{structures}). Only a limited number of
 magnetic structures are compatible with the DO$_{19}$ type symmetry
 \cite{Sticht1989,Brown1990}. Neutron diffraction data was compatible with
 two of them (structures (c) and (d) in Fig. \ref{structures}), but could
 not discriminate between them \cite{Brown1990}. Following Sandratskii and
 K\"ubler \cite{Sandratskii1996}, we focus here on these two magnetic
 structures as well as the two additional ones which are symmetry equivalent
 when SOC is neglected and Sn atoms are assumed non-magnetic (structure (a)
 and (b) in Fig. \ref{structures}).

 One of the most important aspect we are going to discus is the effect of
 the atomic moment approximation (AMA), which is widely used in
 non-collinear magnetic calculations, compared to the exact treatment of
 non-collinearity. We show here that the key role in producing the weak
 ferromagnetic moment in Mn$_3$Sn is not played by the SOC interaction as
 suggested by Sandratskii \textit{et al.}  \cite{Sandratskii1996,
 Sandratskii1998}, but by the precise non-collinear magnetic density on the
 Sn atom which cannot be described within the AMA. In a way it is very
 surprising as the Sn atoms do not carry any magnetic moment on average.

 All the calculations presented in this work were performed using the
 Perdew-Wang LSDA exchange-correlation potential \cite{excPW92}.  We found
 no noticeable differences between test generalized gradient approximation
 (GGA) \cite{pbe} calculations and LSDA ones. Note that, as GGA is not
 formulated for non-collinear spin-densities, we use the gradient of the
 norm of the magnetization density vector as the scalar magnetization
 density entering in the standard formulation.  The details of the setting
 for the L/APW+lo calculations are as follows.  The atomic sphere radii were
 set to 2.5~a.u. for both Mn and Sn atoms. The basis size was setup
 according to $R_\mathrm{MT}K_\mathrm{max}=6.9$ ($R_\mathrm{MT}$ is the
 smallest atomic sphere radius, $K_\mathrm{max}$ is a cutoff for the basis
 functions wave vector). Brillouin zone integrations were performed on a
 $7\times 7\times 7$ mesh (which corresponds to 48 and 64 k-points in the
 irreducible wedge for structures (a), (b) and (c), (d) respectively (see
 Fig.~\ref{structures})).

 We have calculated the electronic structures and total energies of the four
 non-collinear magnetic structures presented in Fig. \ref{structures}. For
 this we used two approaches: (1) AMA, in which the off-diagonal part of the
 spin-density matrix inside the atomic spheres is ignored, and, (2) the
 exact treatment, where no shape approximation is imposed on the
 spin-density and spin-potential matrices. Moreover, each of these
 calculations has been run with and without taking into account SOC.

\begin{figure}
 \includegraphics[width=0.8\linewidth]{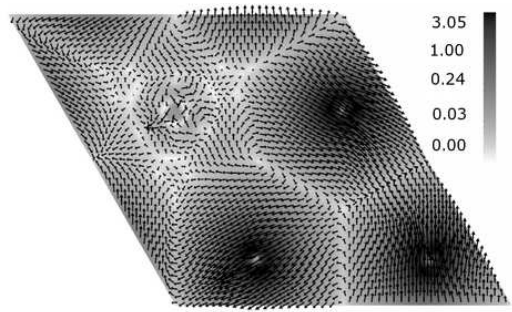}
 \caption{\label{spinden} Calculated spin density for structure (d)
   plotted in the (001) plane cutting three Mn and one Sn atoms
   [$\mu_\mathrm{B}$/a.u.$^3$], using the exact treatment of
   non-collinearity and SOC. An exponential scale is used in order to
   enhance the visibility of regions of low spin density.
 }
 \end{figure}

 Sandratskii and K\"ubler \cite{Sandratskii1996} have already pointed out
 that symmetry analysis is very helpful in understanding the mechanism
 responsible for the advent of weak ferromagnetism in Mn$_3$Sn. Assuming the
 Sn atoms are non-magnetic, they showed that all four structures
 (Fig.~\ref{structures}) are equivalent by symmetry when SOC is
 neglected. Indeed, in the absence of SOC, the spin space can be freely
 globally rotated with respect to the real one.  It must be noted however
 that any magnetization density on the Sn atom would break this equivalency
 irrespective of the inclusion of SOC effects. For example, the (a) and (b)
 structure in Fig.~\ref{structures} appear to be symmetrically related by a
 pure 90$^\circ$ spin-rotation and identity in real space. However, this is
 not an allowed symmetry operation of the magnetic crystal structure as it
 would rotate the spin-density on the Sn site by 90$^\circ$ and there is no
 such four-fold symmetry axis on this atom (see Fig.~\ref{spinden}). Similar
 arguments lead to the fact that the four structures are inequivalent.

 It is clear that the magnetic structure is determined by moments localized
 on Mn atoms. Sn atoms are essentially non-magnetic in the sense that they
 do not carry any average spin moments. Nevertheless, their complicated
 intra-atomic magnetic structure (due to neighboring Mn atoms) has to be
 properly taken into account. We construct the magnetic symmetry group by
 using a two stage algorithm: (1) the symmetry group is determined solely on
 the basis of the localized Mn moments (assuming that the spin-density is
 identically zero within the Sn sites and are thus symmetrical with respect
 to any rotations in spin-space); (2) the quantization axes on the Sn atoms
 are determined in such a way that they preserve the symmetry operations
 that we have just found (using the singular value decomposition technique).
 As a result, the quantization axes of the two Sn atoms are along $+c$ and
 $-c$ directions for both structures (a) and (b), whereas they are in the
 $(ab)$-plane for structures (c) and (d) (i.e. parallel to the Mn2 atoms'
 moments in Fig.~\ref{structures}).  The number of symmetry operations is
 decreased from 24 in the paramagnetic symmetry group to 12 for structures
 (a) and (b), and 8 for (c) and (d). All Mn atoms are equivalent in
 structures (a) and (b), and the triangular arrangement of their moments is
 univocally determined by symmetry.  For structures (c) and (d) the six Mn
 atoms are split into two equivalency classes containing four Mn$1$ and two
 Mn$2$ atoms. The directions of the magnetic moments of Mn$2$ atoms are
 fixed by symmetry, whereas the moments of Mn$1$ atoms can be freely rotated
 in the $(ab)$-plane and have therefore to be optimized during
 self-consistency.

 The results of our total energy calculations are presented in
 Table~\ref{ene}. Our AMA and AMA+SOC calculations confirm the previous
 results of Sandratskii and K\"ubler
 \cite{Sandratskii1996}. Non-relativistic AMA total energies for all
 structures are found in an energy window which is less than 0.1~mRy
 wide. We ascribe the slight lifting of this degeneracy to the fact that, in
 LAPW, AMA cannot be applied to the interstitial region where the
 magnetization must be treated as a continuous field.  SOC shifts down the
 total energies of structures (c) and (d) by about 0.4 mRy with respect to
 structure (a). The total spin and orbital moments calculated within AMA
 (which are zero for structures (a) and (b) due to symmetry) are two to
 three times larger for structure (c) than for structure (d). In both cases
 orbital contributions are about ten times smaller then the spin ones. In
 addition, the orbital moment is parallel or anti-parallel to the spin
 moment for structures (c) or (d) respectively.

 \begingroup
 \squeezetable
 \begin{table}  
 \caption{\label{ene}Total energies (in mRy) of structures (b), (c) and (d)
 relative to the energy of structure (a), and weak ferromagnetic (FM) spin and
 orbital (in parenthesis) moments per unit cell, calculated with different
 approximations. AMA stands for atomic moment approximation, ``Full'' stands
 for exact treatment of the spin-density, M1 means that Mn atoms are treated
 exactly and Sn ones within AMA, M2 means that Mn are treated within AMA and
 Sn exactly. }

 \begin{ruledtabular}
   \begin{tabular}{lccc|r @{} l r @{} l}
                    & \multicolumn{3}{c|}{Total energies [mRy]}& \multicolumn{4}{c}{FM moment [$10^{-2} \mu_B $]} \\

     \hline
                    &   b    & c     &  d    &  \multicolumn{2}{c}{c}   &  \multicolumn{2}{c}{d}    \\
     \hline
        AMA         &  -0.02 & -0.08 & -0.08    &  0.9& & 0.3&  \\
        AMA + SOC     &  -0.06 & -0.42 & -0.42    &  0.8&(0.07)   & 0.4&(-0.04)     \\
        \hline
        Full        &  -0.01 & -2.46 & -2.53    &  6.7 &  & 1.3   &  \\
        Full + SOC    &  -0.01 & -2.90 & -2.98    &  7.5 & (-0.15)   & 1.4 & (0.27)     \\
        \hline
        M1       &  -0.0  & 0.71& 0.71    &  1.6 &  & 1.2  &   \\ 
        M1 + SOC   &  -0.09 & 0.13& 0.13    &  1.5 & (0.27)  & 1.5 & (-0.23)     \\
        M2       &  0.027  & -1.8& -1.76    &  1.6 &  & 1.3 &  \\ 
        M2 + SOC   &  0.52&  -1.57 & -1.59   & 1.5 & (0.27) & 1.8 & (-0.24)   \\
   \end{tabular}
 \end{ruledtabular}
 \vspace*{-\baselineskip}
 \end{table}
 \endgroup

 The most important point we want to address here is the result of our full
 non-collinear calculations. They show that structures (c) and (d) have a
 lower energy (by about 2.5 -- 3 mRy) than structures (a) and (b),
 irrespectively of the inclusion of SOC effects. This is in clear opposition
 to previous conclusions concerning the essential role of SOC in the
 formation of the weak ferromagnetic moment in Mn$_3$Sn
 \cite{Sandratskii1996,Sandratskii1998}. As can be seen from
 Table~\ref{ene}, the effect of spin-lattice coupling is much smaller then
 the effect of switching from AMA to an exact treatment of
 non-collinearity. Similarly to the AMA case, the SOC further decreases the
 total energies of structures (c) and (d) by about 0.4 mRy.  Furthermore, as
 was the case with AMA, the weak ferromagnetic moments for the exact
 calculations are much bigger in the (c) magnetic configuration then in the
 (d) one, but are however about 5 -- 10 times larger than their AMA
 couterparts. In addition, the orientation of the orbital moments changed
 sign relative to the spin moments. For completeness, we want to add that
 the antiferromagnetic collinear structure with moments along $c$ has a
 total energy higher by 34 mRy for AMA and 26 mRy for the exact calculation
 with respect to the energy of structure (a).

 The fact that structures (c) and (d) are stabilized by the exact
 non-collinear treatment, irrespective of SOC effects, emphasizes the
 importance of the off-diagonal terms in the spin-density and potential
 matrices. In order to gain physical insight, we show a map of the
 magnetization density for the (d) configuration in Fig.~\ref{spinden}. The
 relatively simple triangular magnetic structure on the Mn atoms polarizes
 the Sn atoms, resulting in a complicated and interesting magnetization
 density on the Sn atoms. It displays a symmetrical pattern leading to a
 total Sn moment of zero (integrated over the Muffin-Tin volume).  The
 question of the importance of the non-collinearity of the small Sn atom
 spin density to the weak ferromagnetism of Mn$_3$Sn arises. To address
 this, we have carried out calculations in which the Mn and Sn atoms are
 treated within different levels of approximation. These results are
 presented in the bottom part of Table~\ref{ene}, where M1 stands for Mn
 atoms treated exactly and Sn ones within AMA and M2 for the opposite. This
 shows that the M2 case captures the physics of the exact calculation
 (i.e. that the (c) and (d) configurations are stabilized with respect to
 the (a) and (b) ones) whereas the M1 makes the (c) and (d) structures less
 energetically favorable. This means that the intra-atomic non-collinearity
 of Sn atoms plays the essential role in stabilizing the (c) and (d)
 magnetic structures and, therefore, in the weak ferromagnetism
 of Mn$_3$Sn.

 We now consider some atomic properties (see Table~\ref{efg}) which could be
 used experimentally to discriminate between magnetic structures or levels
 of approximation.  As could be expected from their average nature, the
 local atomic magnetic moments are practically the same for both AMA and
 exact treatment and for all structures.  Similarly, the electrical field
 gradients (EFG) \cite{blaha88,blaha2000efg} depend neither on the level of
 approximation used in the calculations (AMA or exact), nor on the magnetic
 configuration (apart, perhaps, for the Sn atom between structures (a), (b)
 and (c), (d)).  Calculated hyperfine fields (HFF) \cite{blugel87} turn out
 to be much more interesting. First, they depend on the methodology (AMA or
 exact) by about 10\% and, second, they are sensitive to the magnetic
 configuration (Table~\ref{efg}). They exhibit a 20\% difference between the
 two inequivalent Mn atoms (bigger by about 20\% on the 4 Mn1 atoms than on
 the 2 Mn2 ones for the (c) configuration and the opposite for the (d) one)
 which could be used to discriminate between magnetic configurations. This
 difference can be ascribed almost entirely on the so-called dipole term
 \cite{blugel87}. It must be noted that HFF are known to be underestimated
 by LDA by about 20\%, but this affects the ``contact term''
 \cite{Novak2003} which does not depend on the details of the magnetic
 configuration and should therefore not influence this difference.

 \begingroup
 \squeezetable
 \begin{table}
 \caption{\label{efg}Electric field gradients (EFG) in $[10^{21}V/m^{2}]$,
 total hyperfine fields (HFF) in kGauss,
 spin (M$_\mathrm{s}$) and orbital (M$_\mathrm{orb}$) moments for Mn and Sn
 atoms in $\mu_\mathrm{B}$. All results include SOC effects. Values for the
 Mn2 atoms (where relevant) are in parentheses.}
 \begin{ruledtabular}
   \begin{tabular}{ccc|ccr @{} l r @{} l}
          & &   &   a  & b  & \multicolumn{2}{c}{c}  &  \multicolumn{2}{c}{d} \\
     \hline
       & Mn &   & 3.03  & 3.03  & \multicolumn{2}{c}{3.03}  & \multicolumn{2}{c}{3.03} \\
     \raisebox{+2ex}[0pt]{M$_\mathrm{s}$}   & Sn &    & 0  & 0  & \multicolumn{2}{c}{0}  & \multicolumn{2}{c}{0} \\
       & Mn &   & 0.04  & 0.04  & \multicolumn{2}{c}{0.03}  & \multicolumn{2}{c}{0.03} \\
     \raisebox{+2ex}[0pt]{M$_\mathrm{orb}$} & Sn &    & 0  & 0  & \multicolumn{2}{c}{0.0015}  & \multicolumn{2}{c}{0.0015} \\
     \hline
        & Mn &  Full & -1.09 & -1.11 & -1.08 & (-1.09) & -1.08 & (-1.07) \\
        & Mn &  AMA  & -1.05 & -1.07 & -1.03 & (-1.04) & -1.04 & (-1.03) \\
     \raisebox{+2ex}[0pt]{EFG} & Sn & Full & -2.12 & -2.11 & \multicolumn{2}{c}{-2.29}        & \multicolumn{2}{c}{-2.29} \\
        & Sn &  AMA  & -2.05 & -2.05 & \multicolumn{2}{c}{-2.17}        & \multicolumn{2}{c}{-2.17} \\
     \hline
          & Mn & Full & 66.7 & 43.1 & 59.5 & (41.2) & 49.1 & (62.5)  \\
     \raisebox{+2ex}[0pt]{HFF} & Mn & AMA  & 60.2 & 38.9 & 53.6 & (40.4) &
          44.5 & (58.5) \\
   \end{tabular}
 \end{ruledtabular}
 \vspace*{-\baselineskip}
 \end{table}
 \endgroup

 To conclude we have presented the results of our {\it ab-initio}
 calculations of the weak ferromagnet Mn$_3$Sn, emphasizing the importance
 of non-collinearity and the short-comings of the atomic moment
 approximation. The comparison between the AMA and exact non-collinear
 calculations shows that the AMA is appropriate for the description of
 atomic properties such as magnetic moments and electrical field gradients.
 However, the AMA leads to the erroneous conclusion that spin-lattice
 coupling is the sole and main mechanism leading to weak ferromagnetism in
 Mn$_3$Sn. Our exact treatment of non-collinearity shows that the weak
 ferromagnetic moment find its origin in the non-collinear intra-atomic
 magnetization density on the Sn atom. To some extent, the fact that such a
 small magnetization density (Fig.~\ref{spinden}) drives a larger decrease
 in energy than spin-orbit coupling effects can be seen as surprising.  The
 magnetic configuration (d) has the lowest energy, but by only 0.08 mRy with
 repect to configuration (c). It is therefore difficult to give a definite
 statement about the most stable magnetic state on that basis. However, the
 total magnetic moment of 0.017 $\mu_\mathrm{B}$/cell (2 formula units per
 unit cell) for structure (d) is in much better agreement with the measured
 moment of 0.01 $\mu_\mathrm{B}$/cell \cite{Tomiyoshi1982} than the five
 times larger moment for structure (c). This suggests that magnetic
 structure (d) is realized by nature.

 We are grateful to the ``EXCITING'' and ``$f$-electron'' European Research
 and Training Networks for providing financial support, and to the Danish
 Centre for Scientific Computing for computational resources.


\end{document}